\documentclass[pra,aps,twocolumn,showpacs,floatfix]{revtex4}
\usepackage{amsfonts,amsmath,bm}

\newcommand{\rl}{\rangle\!\langle}
\newcommand{\rr}{\bm{r}}
\newcommand{\kk}{\bm{k}}

\usepackage{graphicx}

\begin{document}

\title{Intermediate band formation and intraband absorption for electrons in an
inhomogeneous chain of quantum dots} 
\author{Igor Bragar}
\author{Pawe{\l} Machnikowski}
 \email{Pawel.Machnikowski@pwr.wroc.pl}  
\affiliation{Institute of Physics, Wroc{\l}aw University of
Technology, 50-370 Wroc{\l}aw, Poland}

\begin{abstract}
We study the electron states of a chain of
non-identical, vertically stacked quantum dots (QDs).
We discuss
how the pseudo-band formed of the ground states confined in the QDs
disintegrates upon increasing the inhomogeneity of the electron
energies and analyze the impact of localization on the intraband
absorption from the pseudo-band to extended (bulk) states. We describe
also the dependence of the intraband absorption spectrum on the QD size.
\end{abstract}


\maketitle

\section{Introduction} 
\label{sec:intro}

It has been proposed \cite{aroutiounian01} that
inserting a chain  of vertically stacked quantum dots (QDs) in the
intrinsic region of a p-i-n junction solar cell could increase the
photocurrent due to absorption of low energy
(below-bandgap) photons associated with intraband transitions between
the confined states in the QD chain and the bulk continuum.  While
this idea has been gaining 
growing experimental support in the recent years \cite{oshima08,guimard10,okada09,okada11}, it still seems
unclear whether the underlying physics corresponds to the intermediate
band concept \cite{luque97} as the band formation in a quantum dot
chain may be strongly suppressed by the energy inhomogeneity. While
the delocalized nature of the intermediate states (band formation) may
not be essential for the basic idea of providing the possibility for
sequential sub-bandgap absorption, it may affect the rates for the
interband absorption process which will affect the efficiency of the device.

The theoretical modeling of the electron states and optical absorption
in chains and arrays of QDs has so far been limited to infinite,
periodic structures (superlattices)
\cite{deng11,klos09,klos10,tomic08,shao07}. However, actual QD chains
are always finite and usually comprise several to a few tens of
stacked QDs. Even more important is the unavoidable inhomogeneity of
the chain which will obviously modify the essential properties of the
carrier eigenstates.

In this paper, we study the electron states of a chain of
non-identical, vertically stacked quantum dots (QDs).
We use a simple model of a tunnel-coupled chain of
dots containing one electron, which allows us to investigate
the properties of a finite, inhomogeneous system. We study
how the pseudo-band formed of the ground states confined in the QDs
disintegrates upon increasing the inhomogeneity of the electron
energies and discuss the impact of localization on the intraband
absorption from the pseudo-band to extended (bulk) states. We analyze
also the dependence of the intraband absorption spectrum on the QD size.

The paper is organized as follows. In Sec.~\ref{sec:model}, we
describe the model of the system. Next, in
Sec.~\ref{sec:results},  we characterize the eigenstates of
the system and study the intraband absorption. Finally, Sec.~\ref{sec:concl}
concludes the paper. 

\section{Model} 
\label{sec:model}

\begin{figure*}[tb]
\begin{center}
\unitlength1mm
\begin{picture}(128,95)(0,5)
%\begin{picture}(85,64)(0,5)
\put(0,0){\includegraphics[width=40mm]{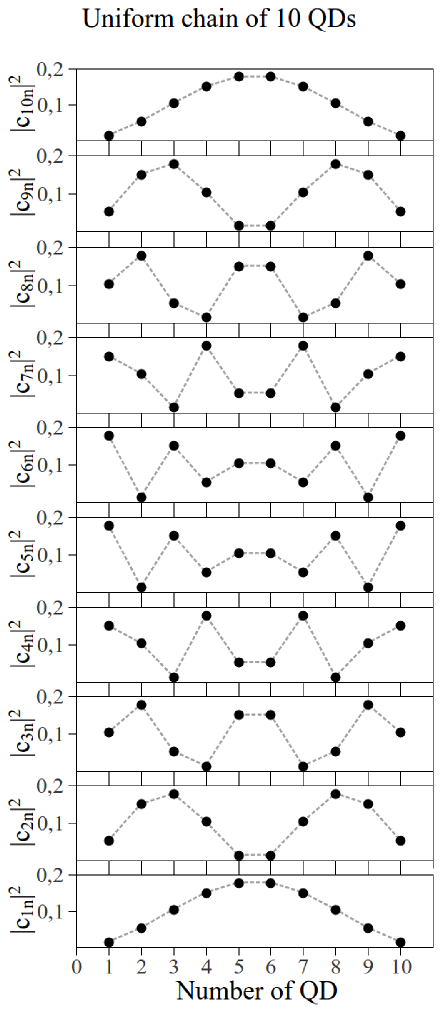}}
\put(44,0){\includegraphics[width=40mm]{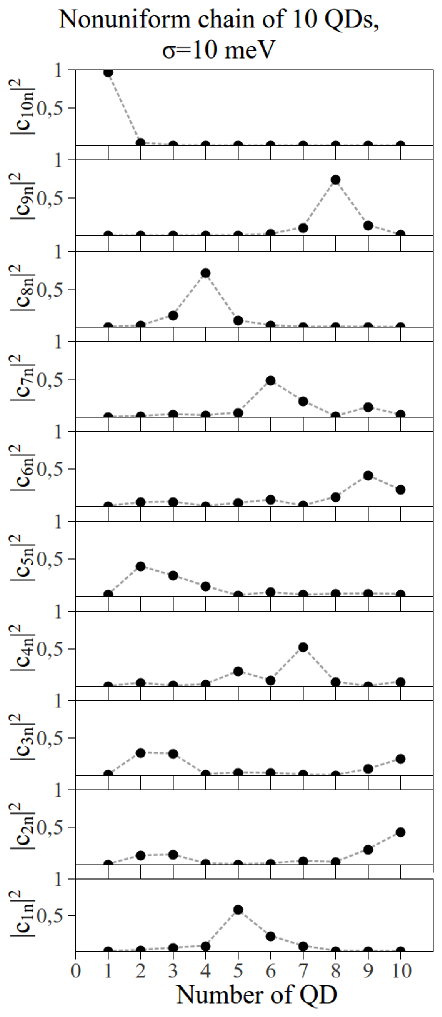}}
\put(88,0){\includegraphics[width=40mm]{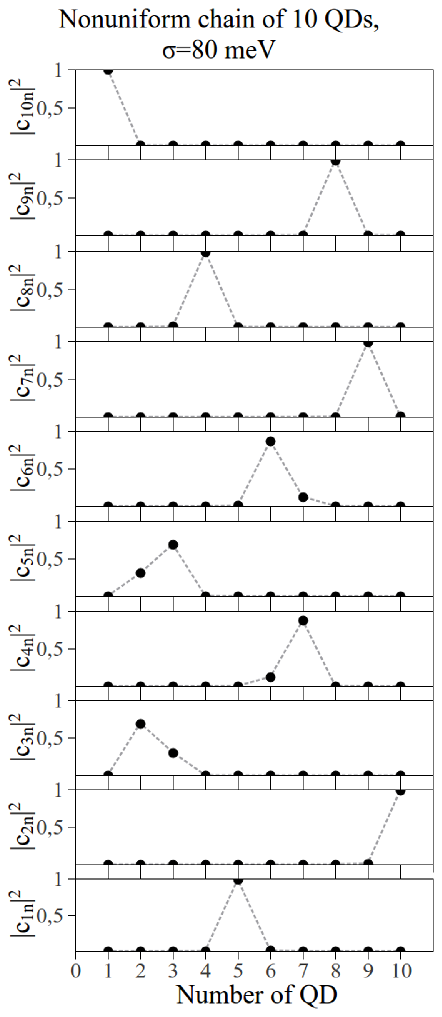}}
\end{picture}
\end{center}
\caption{\label{fig:states}The occupation probabilities for the
electron eigenstates in a chain of 10 identical dots left and in chains
with two different degrees of inhomogeneity (middle and right).}
\end{figure*}

We first consider single-electron states in a chain of QDs stacked
along the growth direction ($z$). Restricting the model to the ground
electron state in each dot and assuming the electron spin to be
fixed, we can write the Hamiltonian of the system in the form
\begin{equation}\label{ham}
H=\sum_{n=1}^{N}\epsilon_{n}|n\rl n|
+t\sum_{n=1}^{N-1} (|n\rl n+1| +\mathrm{h.c.}),
\end{equation}
where $|n\rangle$ denotes the electron state in the $n$th dot with the
wave function $\psi_{n}(\rr)$, $\epsilon_{n}$ is the corresponding energy, $t$ is
the tunnel coupling and we assume the overlap between the
wave functions localized in different dots to be negligible, so
that $\langle n|n'\rangle=\delta_{nn'}$. The usual
inhomogeneity of the QD stack is taken into account by choosing the
energies $\epsilon_{n}$ from the Gaussian distribution with the mean $\overline{E}$
and variance $\sigma^{2}$. The Hamiltonian \eqref{ham} is diagonalized
to yield the single-electron states $|\nu\rangle$ with the wave functions 
$\Psi_{\nu}(\rr)=\sum_{n}c_{\nu,n}\psi_{n}(\rr)$ and energies $E_{\nu}$,
$\nu=1,\ldots,N$.  

\begin{figure}[tb]
\begin{center}
\includegraphics[width=65mm]{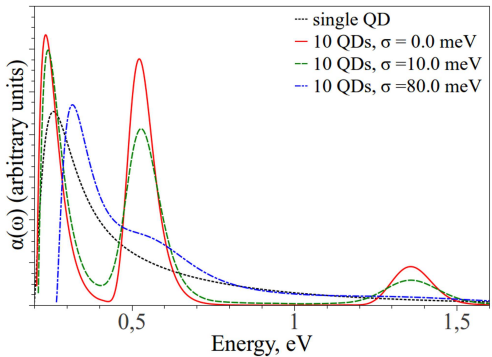}
\end{center}
\caption{\label{fig:dispersion}The intraband absorption spectrum
  for a system initially in the ground state ($\nu=0$) as a
  function of the photon energy $\hbar\omega$ for a single dot and for
QD chains with different degrees of inhomogeneity (disorder), described by the
standard deviation $\sigma$.}
\end{figure}

Next, we calculate the absorption coefficient for intraband dipole
transitions from the initial state $\nu$ to the continuum of bulk
states, which is proportional to
\begin{eqnarray}\label{abs}
\alpha(\omega)\sim
\sum_{\bm{k}} \left| \hat{\bm{\mathcal{E}}}
\cdot \int d^3 r \Psi_{\bm{k}}^{*}(\bm{r})\bm{r}
\Psi_{\nu}(\bm{r}) \right|^2 
\delta\left( \hbar\omega +E_{\nu}-E_{\bm{k}} \right),
\end{eqnarray}
where $\hat{\bm{\mathcal{E}}}$ is the polarization of the incident light,
$\Psi_{\kk}(\rr)$ is the wave function of the final bulk state and
$E_{\kk}$ is its energy.

We assume that the wave functions for the confined states have
the form
\begin{displaymath}
\psi_{n}(\rr)=\frac{1}{\pi^{3/4}ll_{z}^{1/2}}
e^{-\frac{1}{2}\left[ \frac{x^{2}+y^{2}}{l^{2}} +\frac{(z-z_{n})^{2}}{l_{z}^{2}}\right] },
\end{displaymath}
where $z_{n}=nD$ is the position of the $n$th dot ($D$ is the distance
between the dots) and $l,l_{z}$ are the
extensions of the wave function in the $xy$ plane and along $z$,
respectively. 

For the bulk electron states, we assume the simplest approximate model of plane
waves  orthogonalized to
the localized states,
\begin{displaymath}
\Psi_{\kk}(\rr)=N_{\kk}\left[ \frac{1}{\sqrt{V}}e^{i\kk\cdot\rr}
-\sum_{n}\gamma_{\kk n} \psi_{n}(\rr)  \right],
\end{displaymath}
where $N_{\kk}$ is the appropriate normalization constant, $V$ is the
normalization volume 
and the orthogonalization coefficients are given by
\begin{displaymath}
\gamma_{\kk n}=\frac{1}{\sqrt{V}}
\int d^{3}r \psi_{n}^{*}(\rr)e^{i\kk\cdot\rr}.
\end{displaymath}
The corresponding energies are $E_{\kk}=\hbar^{2}k^{2}/(2m^{*})$,
where $m^{*}$ is the effective electron mass.

In the calculations presented in the next section, we consider a chain
of 10 QDs with the inter-dot distance $D=9$~nm. Based on our earlier 
$\bm{k}\cdot\bm{p}$ calculations \cite{gawarecki10}, we set the average energy of
the confined states $\overline{E}=-200$~meV, and the tunnel coupling
$t=-4$ meV. The effective mass is $m^{*}=0.067m_{0}$, where $m_{0}$
is the free electron mass. Each result for a random (disordered) chain has been calculated
for a single realization of the system parameters.

\section{Results} 
\label{sec:results}

In Fig.~\ref{fig:states}, we show the occupation probabilities
$P_{\nu}(n)=|\langle n|\nu\rangle|^{2}$ for a uniform chain (a) and for two
inhomogeneous chains (b,c) with different standard deviations of the
confinement energies $\epsilon_{n}$. For a chain of identical dots, the
probabilities follow the known analytical result 
$P_{\nu}(n)=[1/(N+1)]\sin^{2}[\pi\nu n/(N+1)]$. Thus, an intermediate
quasi-band of delocalized QD states is formed.
As expected,
with increasing inhomogeneity the states tend to localize on
individual dots and the intermediate band disintegrates into localized states.

The localization of the electron states considerably affects the
intraband absorption from the QD states to the bulk continuum. As can
be seen in Fig.~\ref{fig:dispersion} (red solid line), the absorption
probability from the ground state of the chain has
the form of a series of peaks and differs considerably from that
characteristic of a single QD (black dotted line). This results from
the interference 
effect which leads to preferred transitions to states with $k_{z}=2\pi
q/D$, where $q$ is an integer. A maximum appears around the photon
energies 
\begin{displaymath}
\hbar\omega_{q}=|E_{1}|+\frac{2\hbar^{2}\pi^{2}q^{2}}{m^{*}D^{2}},\quad
q=0,1,\ldots,
\end{displaymath}
where new
propagation directions of the final plane wave state become allowed by
energy conservation.  This interference obviously requires that the
initial state is delocalized over the whole chain. As the
inhomogeneity increases the ground state localizes and the absorption
becomes more 
and more similar to the single dot case, apart from a shift, which is
a trivial consequence of shifting the ground state down from the
average energy $\overline{E}$ (dashed and dash-dotted
lines in Fig.~\ref{fig:dispersion}). It is quite remarkable, however, that
the characteristic form of the absorption coefficient survives even
under relatively strong disorder, with the standard deviation of the energies
larger than the tunnel coupling constant.

\begin{figure}[tb]
\begin{center}
\includegraphics[width=65mm]{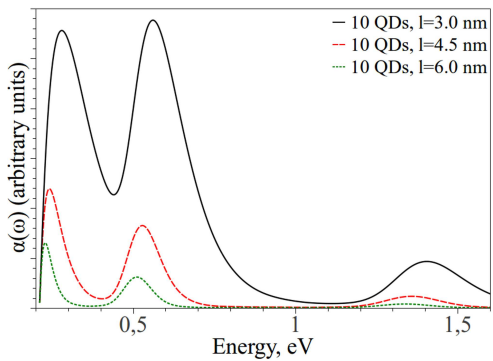}
\end{center}
\caption{\label{fig:l}The dependence of the intraband absorption spectrum
  on the lateral QD confinement size for an inhomogeneous chain with
  $\sigma=10$~meV initially in the ground state.}
\end{figure}

The form of the absorption spectrum depends on the lateral size $l$ of
the confined states in
QDs, as shown in Fig.~\ref{fig:l}. This is related to the increasing
volume of allowed final states in the reciprocal space as the dot size becomes
smaller. Apart from the very strong increase of the magnitude
of the absorption in the whole frequency range
for smaller dots, decreasing the QD size leads to
appearance of non-zero absorption in the gaps between the absorption
peaks, where the absorption is almost null for larger dots. 

\begin{figure}[tb]
\begin{center}
\includegraphics[width=65mm]{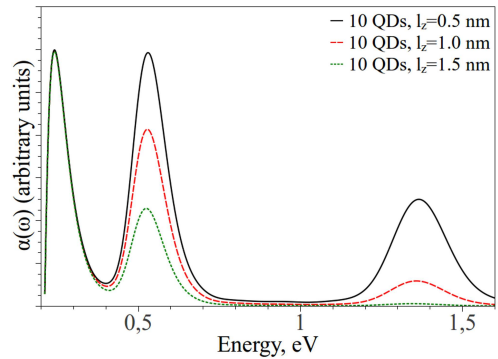}
\end{center}
\caption{\label{fig:lz}The dependence of the intraband absorption spectrum
  on the QD confinement size in the growth direction for an inhomogeneous chain with
  $\sigma=10$~meV initially in the ground state. }
\end{figure}

The dependence of the absorption spectrum on the dot height (the
confinement size
in the growth direction, $l_{z}$) is shown in Fig.~\ref{fig:lz}. While
changing the QD height affects the first peak very weakly, the
dependence becomes important for higher peaks that disappear as
$l_{z}$ becomes larger. This results from the fact that the higher peaks
are built upon the tail of the single dot absorption
(cf. Fig.~\ref{fig:dispersion}), which is due to transitions to
large momentum states propagating in the strongest confinement
direction. When the single dot states become more weakly localized in
this direction the momentum selection rules restrict these transitions
and the tail vanishes, which also leads to disappearance of the higher
absorption peaks.

\section{Conclusions}
\label{sec:concl}

We have studied the properties of electron states in a finite, inhomogeneous chain of
quantum dots and the intraband absorption spectrum related to
transitions from the lowest confined state in the QD chain to the bulk
continuum states. We have shown that even for a relatively short chain
of 10 dots, the intraband absorption spectrum is dominated by
interference effects and differs considerably from a single dot
spectrum. This effect is remarkably stable against the random variation
of the energy levels in the individual dots and persists even when the
standard deviation of the latter is a few times larger than the
magnitude of the tunnel coupling.
 For dots with typical confinement sizes, the spectrum has
the form of broad peaks separated by gaps where the system shows
almost no absorption. The gaps can be closed by reducing the lateral
dot size. On the other hand, increasing the size in the growth
direction considerably reduces the magnitude of absorption peaks at
higher frequencies.

\textbf{Acknowledgment:} This work was supported by the Foundation for
Polish Science under the TEAM programme, co-financed by the European
Regional Development Fund. 

%\bibliographystyle{prsty}
%\bibliography{abbr,quantum}

\end{document}